# Formation of $Mn^{2+}$ in $La_{2/3}Ca_{1/3}MnO_3$ Thin Films due to Air Exposure

S. Valencia*, A. Gaupp and W. Gudat
BESSY, Albert-Einstein-Str. 15, D-12489, Berlin, Germany

Ll. Abad, Ll. Balcells, A. Cavallaro and B. Martínez
Institut de Ciència de Materials de Barcelona, Campus de la UAB, E- 08193 Bellaterra, Spain

F. J. Palomares
Instituto de Ciencia de Materiales de Madrid, CSIC, Cantoblanco, E-28049 Madrid, Spain

**Abstract**

We report on the chemical stability of $La_{2/3}Ca_{1/3}MnO_3$ thin films. X-ray absorption spectroscopy at the Mn L-edge and O K-edge makes evident deviations from the nominally expected (2/3-1/3) $Mn^{3+}/Mn^{4+}$ ratio after the growth of thin films on $LaAlO_3$ substrates. As-grown thin films, exhibiting Curie temperature, $T_C$, well below that of the LCMO bulk material, develop an unexpected $Mn^{2+}$ contribution after a few days of air exposure which increases with time. Moreover, a reduction of the saturation magnetization, $M_S$, is also detected. The similarity of the results obtained by electron yield and fluorescence yield demonstrates that the location of the Mn valence anomalies are not confined to a narrow surface region of the film but can extend throughout the film thickness in case of granular films. High temperature annealing not only improves the magnetic and transport properties of such as-grown films but also recovers the expected 2/3-1/3 $Mn^{3+}/Mn^{4+}$ ratio, which thereafter is stable to air exposure. Similar results on $La_{2/3}Ca_{1/3}MnO_3$ films grown on $SrTiO_3$ and $NdGaO_3$ substrates demonstrate that there is no direct relation between the observed Mn valence instability and the strain state of the films due to their lattice mismatch with the substrate. A mechanism for the formation of $Mn^{2+}$ ions formation is discussed.

*Corresponding author, e-mail: Valencia@bessy.de

## I. Introduction

The physics of rare-earth manganese perovskites has extensively been studied to understand the origin of the strong correlation between their magnetic and transport properties. As outcome of these investigations, a complex scenario has been suggested where in addition to the double exchange theory [1-3] and the Jahn Teller distortions of the lattice [4], charge and orbital degrees of freedom as well as a natural tendency of the material towards phase separation is taken into account [5-7] in order to describe the observed behaviour.

The implementation of magneto-electronic thin film devices based on these materials is still a serious problem, since their transport and magnetic properties exhibit a strong dependence on the preparation methods and conditions. The origin of which is not well established yet. Structural mismatch with the substrates [8], inhomogeneities located at interfaces [9], surface segregation [10-12] and oxygen depletion [13], are only some of the possible explanations offered so far.

Not much attention has been paid to the manganese valence stability. Brousard et al. reported on the stability of the manganese valence to some minutes of air exposure [14] for $La_{2/3}Ca_{1/3}MnO_3$ (LCMO) thin films on $SrTiO_3$ (STO) substrates. Early investigations by Hundley et al. [15] suggested the presence of $Mn^{2+}$ in polycrystalline $La_{1-x}Ca_xMnO_{3+\delta}$ samples. More recently de Jong et al. [16], found direct experimental evidence for $Mn^{2+}$ at the surface of as-grown epitaxial $La_{2/3}Sr_{1/3}MnO_3$ (LSMO) thin films on STO substrate by means of X-ray absorption spectroscopy (XAS). They suggested the possibility that the $Mn^{2+}$ corresponding structure in the spectrum was earlier not correctly interpreted due to a wrong identification with a change in the nominal $Mn^{3+}/Mn^{4+}$ valence ratio.

In this paper we present a systematic XAS investigation of Mn valence changes on a set of LCMO thin films grown on LaAlO$_3$ (LAO) (001)-oriented (cubic notation) substrates. We used total electron yield (TEY) probing a 4-5nm thin surface region as well as fluorescence yield (FY) detection known to be spectrally much more bulk sensitive dependent. Spectra were measured at the Mn L-edge and at the C and O K-edges. The former directly probe the unoccupied Mn 3d states via 2p ➔ 3d transitions. The latter probe the unoccupied O 2p states via the O 1s ➔ 2p dipolar transitions to give information on the Mn 3d occupancy and hence on the Mn valence, due to the hybridization between the O 2p and Mn 3d orbitals [17].

## II. Experiment

About two months before the XAS synchrotron radiation experiments a set of three LCMO films were simultaneously grown on LAO substrates by means of rf magnetron sputtering. During deposition the substrate temperature was kept at 800 °C. The pressure was 330 mTorr (Ar- 20% O$_2$). The films were in-situ annealed at the same temperature at an oxygen pressure of 350 Torr for 1 hour. Afterwards, they were cooled down to room temperature at 15°C/min rate within the same oxygen atmosphere.

After finishing the growth process samples were kept under different atmospheric conditions. Sample *Ia* was kept in ambient air. Sample *Ib* was immediately held under vacuum in a dry box (desiccator) to minimize air exposure; residual interaction with the atmosphere of almost one day must be considered due to manipulation of the film. The third one (*Ic*) was annealed in air for two hours at T=1000ºC, with heating and cooling ramps of 5ºC/min, and thereafter also kept in air for about two months.

In order to compare the effects of exposure to different conditions over time, a second set of LCMO thin films (*IIa*, *IIb* and *IIc*) was prepared ten days before the synchrotron experiments. As a reference sample for the XAS spectra, a piece of the LCMO bulk target used for the growth of the films, with proper magnetic and structural properties of the 2/3-1/3 composition, was also measured.

The thicknesses of the thin film samples as well as their roughness were deduced from grazing incidence x-ray reflectometry (XRR). The out-of-plane cell parameter *c* was obtained by means of x-ray diffraction (XRD) experiments at the (004) reflection. The magnetotransport properties were measured in a four-probe configuration with a Quantum Design physical properties measurement system (PPMS) in the temperature range 10–300 K and with a maximum field of 30 kOe applied perpendicular to the plane of the samples. Contacts were made by attaching platinum wires to the samples with silver paste.

Magnetic hysteresis at T=10 K and magnetization curves as function of temperature were measured for an external in plane magnetic field of H=5000 Oe with a superconducting quantum interference device magnetometer (Quantum Design). The Curie temperature was determined by the interception of the high temperature magnetization slope with the axis at zero magnetization.

The XAS experiments were performed at the undulator beamline UE56-1-PGM-1 of the synchrotron radiation source BESSY [18]. The spectral resolution at the Mn 2p and C and O 1s edges was ca. $E/\Delta E=5000$ and the degree of polarization was set to circular (right helicity and $P_{circ}=0.90\pm0.03$). We used the BESSY ultra-high vacuum polarimeter chamber [19], which allows the simultaneous measurement of TEY, FY and reflectivity. For the TEY detection the photo excited drainage current of the sample was recorded while the sample was kept at a potential of -95V with respect to the chamber. The fluorescence detector, a GaAsP photodiode was placed aside as close as possible to the sample. Nearby drainage electrodes were kept at a potential of +400 V. The angle of grazing incidence was fixed to $\phi_i = 40^o$, which is known from previous investigations to avoid saturation effects for the TEY data [20].

### III. Results

### IIIa. Structural and magnetic properties

The structural and magnetic properties of the samples kept in air (*Ia/IIa* and *Ic/IIc*) were measured ten days after the growth process (which includes the annealing in case of the *Ic/IIc* films). Those kept in vacuum (*Ib/IIb*) were measured after finishing the synchrotron experiments to minimize air exposure before the XAS measurements.

A thickness *t* of 21±1 nm and a root mean squared (*rms*) surface roughness of 4±1 Å was obtained by fitting the XRR curves for the samples of set *I*. The thickness of samples of set *II* was *t*=10±3 nm with indications of larger surface *rms* roughness. Samples *Ia/Ib* and *IIa/IIb* have larger out-of-plane cell parameter *c*=3,945±5 Å and *c*=3,950±5 Å, respectively, when compared to the LCMO bulk (3.860 Å) due to the compressive strain induced by the in-plane mismatch with the substrate (3.79 Å), in agreement with previous results reported for films of similar thickness [21]. The annealed samples known to release part of the structural strain by an increase of the in-plane cell parameter relax to *c*=3.883±2 Å and *c*=3.930±2 Å for *I*c and *II*c, respectively, approaching the bulk value.

The magnetic measurements show for all of the non-annealed samples of set *I* (see Fig. 1) a similar reduction of $T_c$ with respect to the bulk value (270 K). Nevertheless, large differences appear when comparing their saturation magnetization $M_s$. While the *Ib* sample kept in vacuum has a $M_s$ value slightly different from that of bulk material (580 emu/cm$^3$), the *Ia* film (air exposed) shows a significant reduction. Its origin, as it will be shown later is related to the period of time during which *Ia* sample was exposed to air. This reduction is even larger for samples of set *II* due to their smaller thickness. On the other hand the transport properties of these samples indicate that they have a strong granular character [22]. After the high temperature annealing treatment all the samples exhibit an epitaxial-like structure and a substantial improvement of their magnetic and transport properties approaching those of the bulk material. For instance, sample *Ic* exhibits $M_s \approx 580$ emu/cm$^3$ and $T_c \approx 270$ K very close to bulk value. But it is to be noted that sample *IIc* still presents slightly depressed values of $T_C$ and $M_S$ with a smaller degree of relaxation of in-plane strain. The latter can be inferred from the smaller enhancement of the *c* cell parameter. A complete report of the structural, magnetic and transport properties of these samples will be published elsewhere [22].

**IIIb. XAS**

Figure 2 shows the TEY Mn L- edge spectra for the 3 samples of set *I*. In order to facilitate comparison between them, the spectra have been normalized at the energy (643.2 eV) where the maximum intensity is measured for the sample *Ic*. However, other normalization criteria have been shown not to influence qualitatively our conclusions.

The spectrum of the annealed sample *Ic* is very similar to that of the LCMO bulk sample, even after a long period of air exposure. We take this similarity as indication that the annealed films have the same $Mn^{3+}/Mn^{4+}$ ratio as the bulk and thus a hole doping corresponding to the nominal Ca doping of x=0.33. Clear differences appear depending on the history of the thin films. The sample kept in the desiccator (*Ib*) and even more the one kept in air (*Ia*) exhibit large differences at the low energy side of the $L_3$ and $L_2$ peaks. Inset of fig. 2 shows the difference between the spectra of samples *Ia* and *Ic* and that of *Ib* and *Ic*. For comparison the theoretical $Mn^{2+}$ ($3d^5$) calculated spectrum [23] for a $Mn^{2+}$ ion in a crystal field splitting of 0.5 eV and tetrahedral symmetry is also shown . Good agreement with respect to peak positions as well as to the spectral shape is found between this theoretical spectrum and the *Ia-Ic* difference spectrum, showing that the extra features for the as-grown sample kept in air are only due to the presence of $Mn^{2+}$.

For the as-grown sample kept under vacuum conditions a unique $Mn^{2+}$ extra component cannot be deduced from its difference with respect to the *Ic* spectrum. Although *Ib-Ic* shows characteristic spectral features originating from a $Mn^{2+}$ presence such as the shoulder at E≈640.6 eV and the small peak at E≈ 644eV, much larger differences appear on the higher energy side of the $L_3$ and $L_2$ peaks. In order to correctly subtract the $Mn^{2+}$ spectral contribution from it and see the origin of such differences, *Ib-Ic* has been recalculated by normalizing the *Ic* spectrum to 0.97*Ic* to avoid negative contributions to the difference as shown in figure 3. The correct scaling of the theoretical $Mn^{2+}$ spectrum for subtraction has been decided to be that which largely reduces the intensity at the pre-edge region of the $L_3$ peak, where the major $Mn^{2+}$ contributions are. A clear shift towards higher energies is observed for both L edges of the resulting curve when compared to the $Mn^{2+}$ theoretical spectrum. Since $Mn^{3+}$ and $Mn^{4+}$ have larger binding energies, we interpret these results as the presence of regions with an excess of $Mn^{3+}$ or/and $Mn^{4+}$ and thus with an improper $Mn^{3+}/Mn^{4+}$ ratio. Its comparison with the spectrum corresponding to the bulk sample points mainly to an excess of $Mn^{3+}$.

The $Mn^{2+}$ presence is complementary supported by the data obtained at the O K edge by TEY. The origin of the broad peaks at 536eV and 544eV are attributed to bands of La 5d and Mn4 sp and La 6sp character, respectively [17]. The peak at 530 eV is due to dipole transitions from O 1s to O 2p orbitals which are hybridized with those of unoccupied of Mn 3d. Thus, the intensity of this peak represents an indirect measure of the Mn 3d level occupancy. Fig. 4 shows the measured spectra for the 3 samples of set *I* normalized to the same area (assuming similar numbers of free states and similar oscillator strength). The size of the low energy peak of sample *Ic* is similar to that measured for the bulk sample [24] indicating a similar occupancy and thus a similar $Mn^{3+}/Mn^{4+}$ valence ratio. A decrease of this peak is observed for the *Ia sample*. This indicates a reduction of the unoccupied states at the Mn 3d level, reflecting the presence of more electrons than expected for nominal hole doping.

In order to determine whether this $Mn^{2+}$ emission is restricted to the outermost layers of the films, the TEY spectra are compared with those obtained by means of FY detection. Our fluorescence yield data show similar results at both edges (Mn and O) revealing that those "anomalies" extend through the thickness of the film. Fig. 5 shows a comparison between the TEY and FY data for the *Ia* sample at the Mn L-edge. Differences with respect to the $L_3/L_2$ ratio can be noticed after normalization, most likely due to saturation effects [25] for the FY data at the selected angle of incidence.

Fig. 6 shows the Mn L-edge for the films of set *II* grown ten days before the experiments. Spectra of samples *IIb* and *IIc* are very similar to those of samples *I*b and *I*c, respectively (see Fig. 2). On the other hand, clear differences can be appreciated between the spectra of samples *Ia* and *IIa*. Some contribution of the $Mn^{2+}$ peak in the spectrum of sample *IIa* may be inferred, but this contribution is not as strong as that detected for sample *Ia*. In agreement with this observation, the Mn 3d-related peak of the O K-edge spectrum for the sample kept in air (*II*a) shows corresponding differences with respect to that of the annealed one (Fig. 7).

From the comparison of data corresponding to samples of series *I* and *II* the temporal evolution of the Mn valence state for an as-grown LCMO/LAO thin film exposed to ambient air can be deduced. Unfortunately we cannot perform *in situ* XAS measurements which preclude determination of the actual Mn valence of as-grown samples prior to venting the evaporation chamber. Nevertheless, the similarity between the spectra of *Ib and IIb* films, in spite of their difference in age indicates that the vacuum environment of the desiccator stabilizes the Mn valence avoiding the degradation of the films. This allows us to suggest that the Mn L-edge spectrum of as-grown samples prior to any contact with the air atmosphere will look very similar to those obtained from the *Ib* and *IIb* thin films. A $Mn^{3+}/Mn^{4+}$ valence ratio other than that expected of the nominal composition occurs in the growth process. The Mn valence of the atoms of those regions is particularly unstable in air and after ten days a fraction of it has chemically been reduced to $Mn^{2+}$ as deduced from the spectra of sample *IIa*. Samples kept in vacuum, for which a residual interaction of almost one day should be kept in mind, already indicate such a conversion. Finally, for longer periods exposed to air as is the case of film *Ia* (2 months), a complete reduction of this Mn towards $Mn^{2+}$ takes place.

Annealing the as-grown samples in air at 1000°C recovers not only bulk-like properties (cell parameter, $M_s$, $T_c$), but also the nominal $Mn^{3+}/Mn^{4+}$ ratio, since it has been observed that their XAS spectra resemble that of the bulk sample. Films with the correct stoichiometry are stable towards air exposure at room temperature, at least for a two months period.

In order to check whether the developed $Mn^{2+}$ component had any detectable effect on the structure and/or magnetic properties, sample *Ia* was again characterized after two months of air exposure. Neither changes of the out-of plane cell parameter nor in the transition temperature Tc were observed but, as shown in Fig. 1, a further reduction of $M_s$ was found [22].

## IV. Discussion

We now discuss the mechanism of the formation of the $Mn^{2+}$ ions in LCMO films. We found that a stabilization of the Mn valence is accomplished after high temperature annealing in air. We have demonstrated that vacuum as the one in the desiccator, stabilizes the Mn valence of the films even when they do not present the expected 2/3-1/3 ratio, whereas exposure to air contributes to $Mn^{2+}$ formation. De Jong et al. [16] demonstrated that annealing LSMO films in vacuum which removes oxygen [26], increases the relative height of the $Mn^{2+}$ component of the Mn L-edge spectrum, whereas annealing in a controlled oxygen atmosphere reduces it.

From these results a picture where the $Mn^{3+}/Mn^{4+}$ valence balance is unstable at air exposure due to oxygen deficiency in the films after the growth, can be speculated on. In order to compensate the lack of oxygen and maintain the electrical neutrality, an excess of $Mn^{3+}$ should therefore be present in as-grown samples. This ion can be chemically reduced to $Mn^{2+}$ due to interaction with air or oxidized to $Mn^{4+}$ by means of an annealing process recovering the nominal stoichiometry [13, 26].

In order to explain the presence of $Mn^{2+}$ not only for regions close to the surface but also within the films we should mention that this oxygen deficiency can occur not only in regions close to the films free surface but also at grain surfaces in granular thin films. Therefore, their influence should be strongly enhanced in films with a marked granular character as those of the present study, as suggested by the magneto-transport data [22]. This is in agreement with similar measurements [24] from a set of LCMO films grown on top of $SrTiO_3$ and $NdGaO_3$ (NGO) substrates (with 1% and almost zero lattice mismatch, respectively). A clear $Mn^{2+}$ signature was observed for films grown on NGO, known by atomic force microscopy images of the surface to present a strong granularity. On the other hand, in good quality epitaxial thin films as those grown on STO substrates [9] the effect of imbalance of the $Mn^{3+}/Mn^{4+}$ composition and oxygen deficiency is likely to be restricted to an area close to the surface [16]. This has in fact been corroborated by Hall effect measurements in as-grown and annealed LCMO/STO epitaxial thin films [27].

Interesting enough, the existence of a region close to the free surface with presence of $Mn^{2+}$ and therefore with an imbalance of the $Mn^{3+}/Mn^{4+}$ composition can explain why surface resistance is larger in as-grown than in annealed films as observed in AFM current sensing measurements in LCMO/LAO samples [28]. These results also offer some clues to better understand effects such as the reduction of spin polarization close to the film surface and its faster decrease with temperature [29].

The $Mn^{2+}$ ion formation is not exclusively observed on $La_{2/3}Ca_{1/3}MnO_3$ thin films. Similar results were also detected in $La_{0.5}Ca_{0.5}MnO_3$ samples [30]. Furthermore, as commented by de Jong et al. the presence of $Mn^{2+}$ might have been previously incorrectly related to an increase of $Mn^{4+}$ as well as to a change of the crystal field strength in both $La_{1-x}Ca_xMnO_3$ and $La_{1-x}Sr_xMnO_3$ films. Thus it seems to be a general feature occurring on Lanthanum-Manganese perovskites.

A reducing media present in the air must be at the origin of the Mn reduction. R. Cracium et al. have demonstrated that in $MnO_x$-YSZ catalytic materials, $Mn^{3+}$ can provide sites for CO adsorption and supply oxygen from its oxide structure for oxidation, leading to $Mn^{2+}$, $CO_2$, or $CO_3^{2-}$ formation [31]. In our case presence of Carbon has been detected on all samples by means of XAS (TEY and FY) measurements at the C K-edge. Thus a similar mechanism might be at the origin of the $Mn^{2+}$ formation in our films. We suspect that after the growth process those regions with the observed suboptimal $Mn^{3+}/Mn^{4+}$ may be metastable and vulnerable to the atmospheric reduction. Further investigations are, however, needed in this direction in order to clarify the Mn valence balance control in free surfaces and interfaces since they do play a very important role on the implementation of oxide-based thin films magneto-resistive devices.

## V. Conclusions

Our results strongly suggest that as-grown LCMO thin films prepared by magnetron sputtering suffer from a $Mn^{3+}/Mn^{4+}$ valence balance instability tentatively related to be due to a oxygen deficiency in the films. Exposure to ambient atmosphere of these LCMO as-grown samples, give rise to the appearance of $Mn^{2+}$ ions in a region close to the free surface of the films. This effect is much more pronounced for granular films than for epitaxial ones, as demonstrated by our XAS experimental results. In fact, for granular films we find $Mn^{2+}$ spectral signatures throughout the film thickness. The nature of the mechanism leading to the formation of the $Mn^{2+}$ is possibly related to an initial deviation from the nominal $Mn^{3+}/Mn^{4+}$ ratio caused by an oxygen deficiency. Concomitant with this, as-grown samples exhibit reduced magneto-transport properties, in comparison with bulk material. This reduction increases after a long lasting air exposure. In contrast, samples kept under vacuum conditions (dry box) do not show this aging process, provided they are not exposed to air. Only samples subjected to a high temperature (1000°C) annealing process, even if they were subsequently exposed to the ambient atmosphere, exhibit a clear improvement of their magneto-transport properties, namely increase of $T_C$ and $M_S$ and decrease in resistivity. Our studies clearly indicate that this effect of $Mn^{3+}/Mn^{4+}$ valence imbalance should be mostly confined to a narrow region close to the film free surface in epitaxial thin films. But it is strongly reinforced in granular films, as those of the present study, due to the enlarged free surface of the grains, ad deduced from the fluoresce yield XAS data. The existence of a region with reduced magneto-transport properties close to the surface can be responsible for the enhancement of the resistivity and the stronger degradation of the spin polarization observed close to the surface in manganite thin films. We strongly believe that our results are of major relevance for the fabrication of oxide-based magneto-electronic devices that require in most cases the use of thin films for their implementation, and clearly point towards the necessity of a better control of the growth process of thin film oxides and a better knowledge of the chemistry of free surfaces and interfaces.

**Acknowledgments**

One of the authors (S. Valencia) would like to thank Dr. L. Soriano for valuable discussions. L. Abad, LL. Balcells and B. Martinez would like also to acknowledge financial support from the MCyT (Spain) and FEDER (EC) project MAT2003-04161.

**Figure captions**

Figure 1. Magnetic hysteresis cycles (right) measured at T=10K. The results for the *Ia* sample measured after the XAS experiments are shown with a continuous line. Inset: Magnetization curve in function of temperature measured with H=5000 Oe

Figure 2. Manganese L-edge XAS spectra measured by TEY for the samples *Ia*, *Ib* and *Ic* grown 2 months before experiments. The spectrum for a bulk reference sample (red continuous line) is shown for comparison. The inset shows the differences between *Ia*, *Ib* and *Ic* spectra. The theoretical spectrum for $Mn^{2+}$ is also plotted (continuous line). The presence of $Mn^{2+}$ is evident in the as-grown sample expose to air

Figure 3. Comparison between *Ib-0.97Ic* (full circles), the theoretical $Mn^{2+}$ spectrum (continuous line) and its difference (open circles). A clear shift towards higher energies is present for the latter indicating the presence of regions with a $Mn^{3+}/Mn^{4+}$ ratio different than the nominally expected for the sample kept under vacuum conditions.

Figure 4. Oxygen K-edge XAS spectra measured by TEY for the samples Ia, Ib and Ic, normalized to the area.

Figure 5. Comparison between the Mn L-edge spectra obtained by means of TEY (filled dots) and FY (open dots) for the sample kept in air *Ia*.

Figure 6. Manganese L-edge XAS spectra measured by TEY for the samples *IIa*, *IIb* and *IIc* grown 10 days before experiments. The spectrum for a bulk reference sample (red continuous line) is shown for comparison. The inset shows the differences between *IIa*, *IIb* and *IIc*

spectra. For comparison the theoretical spectra for $Mn^{2+}$ is also plotted. Some $Mn^{2+}$ has occurred after of exposure to air.

Figure 7. Oxygen K-edge XAS spectra measured by TEY for the samples Ia, Ib and Ic, normalized to the area.

**Figure 1**

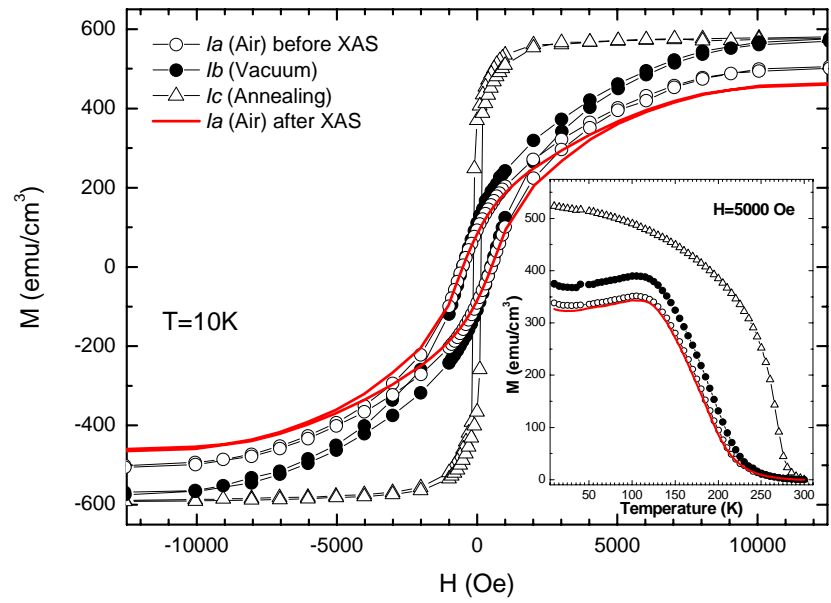

**Figure 2**

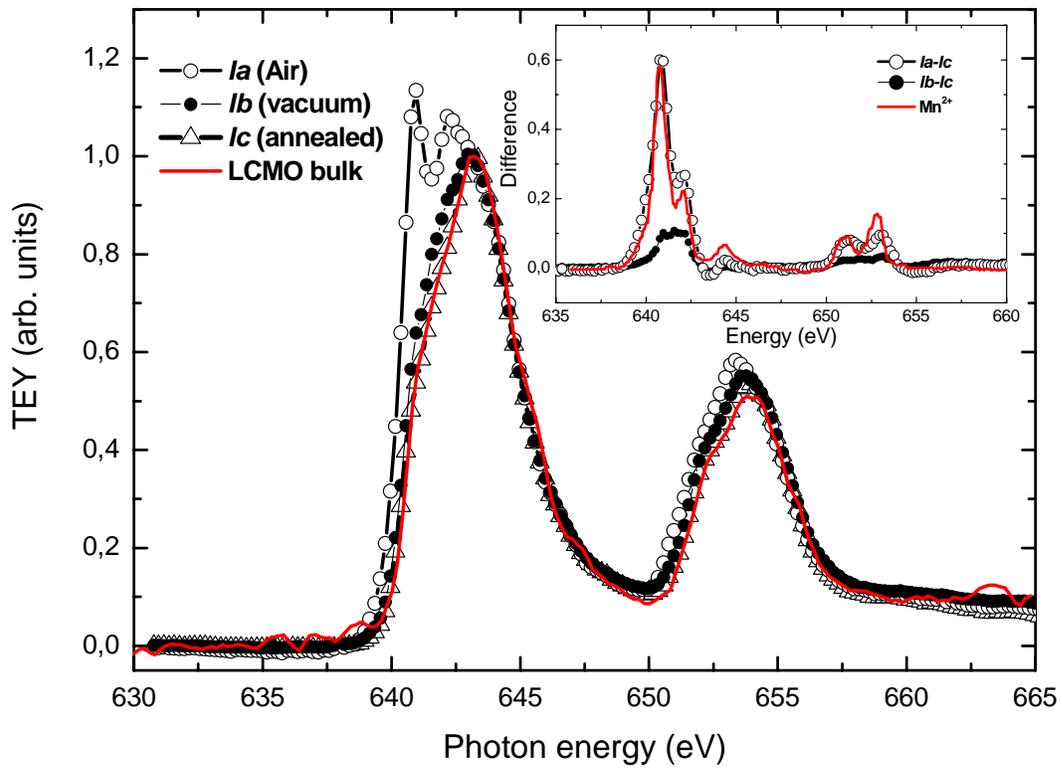

**Figure 3**

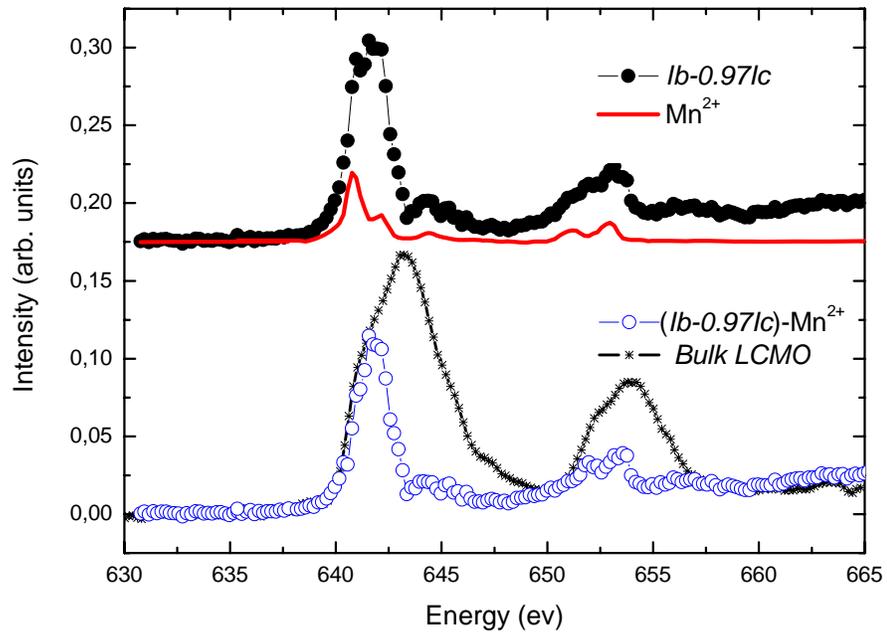

**Figure 4**

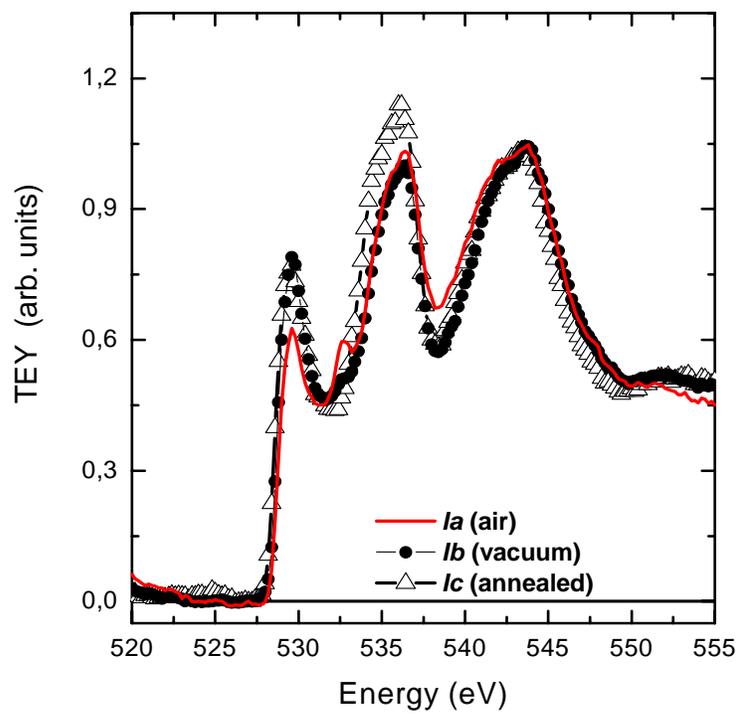

**Figure 5**

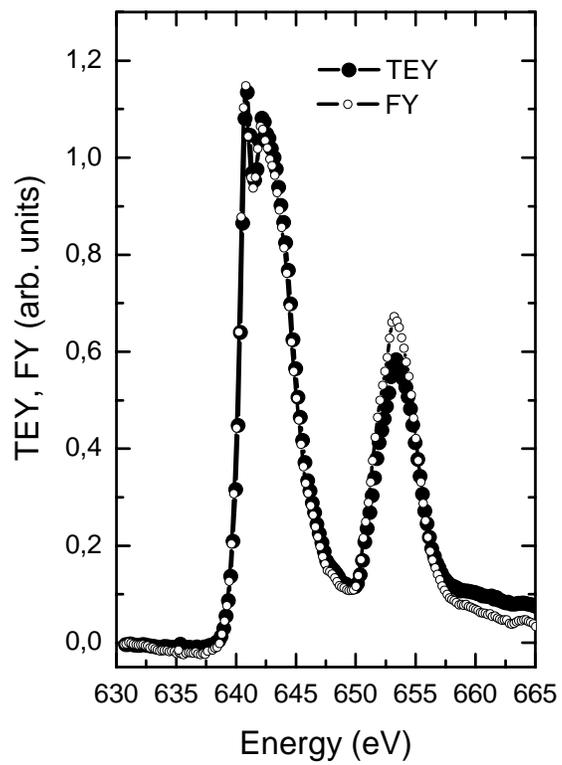

**Figure 6**

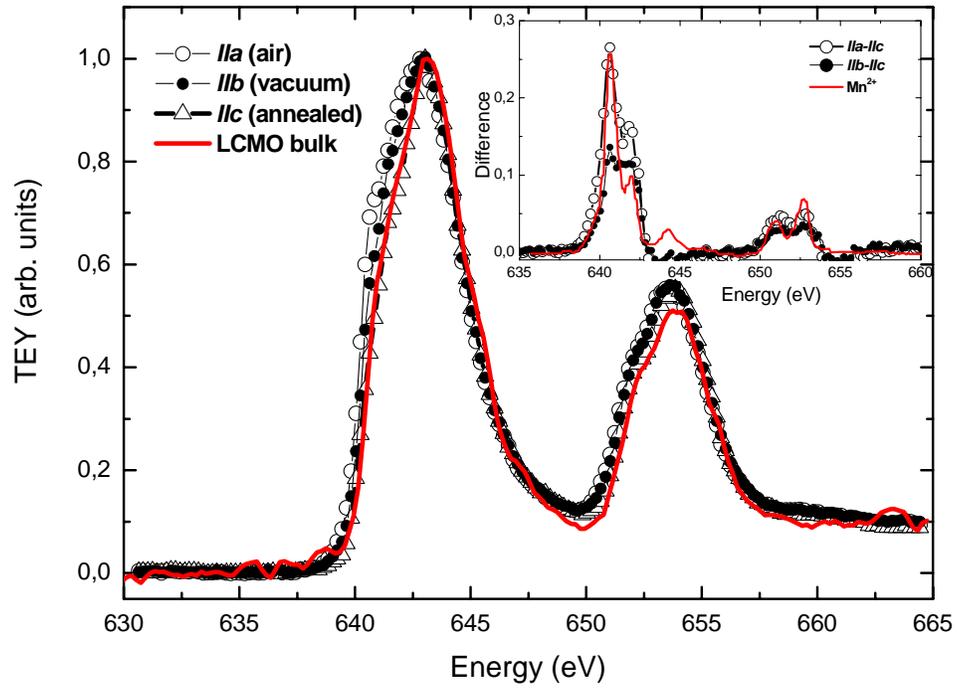

**Figure 7**

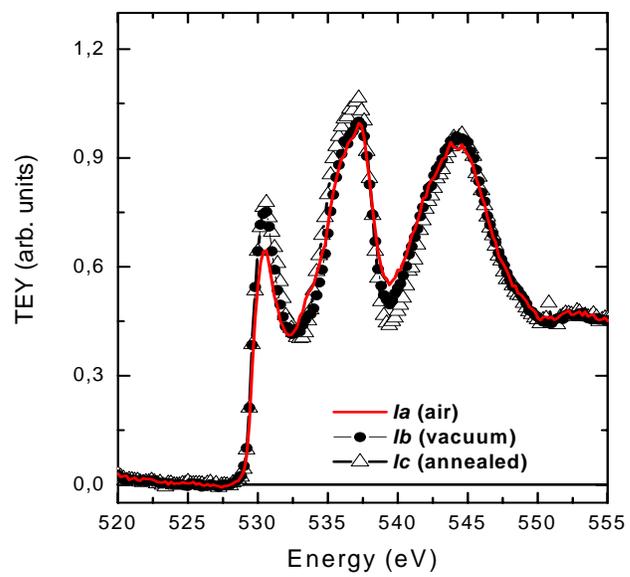